\documentclass[12pt,preprint]{aastex}

\lefthead{Escala}
\righthead{Evidence in Favor of Direct Massive Black Hole Formation.}

\begin{document}

\title{
Flat Galactic Rotation Curves 
Interpreted as Evidence for the Mach's Principle  }

\author{Andr\'es Escala}
\affil{Departamento de Astronom\'{\i}a, Universidad de Chile, Casilla 36-D, Santiago, Chile.}
\affil{aescala@das.uchile.cl}

\begin{abstract}


{We explore the possibility of a Machian interpretation for the phenomena associated with asymptotically flat rotation curves in disk galaxies, providing an alternative to both the Lambda-Cold Dark Matter ($\Lambda$CDM) paradigm and MOdified Newtonian Dynamics (MOND). We argue that the MOND acceleration scale $a_0$ is likely not a new fundamental constant, but rather a manifestation of the cosmological scales of the observable Universe ($a_0 \sim c^2/R_u \sim  c^4/GM_u \sim c/t_u$). We rewrite the Baryonic Tully-Fisher Relation (BTFR) in a cosmological format, $\frac{M_{\rm bar}}{M_{u}} = \left( \frac{V}{c} \right)^4$, suggesting that the transition to the non-Newtonian regime occurs when local accelerations become small enough to feel the global cosmic boundary, and showing that any Machian interpretation must fulfill a relation of this form. Furthermore, this framework naturally predicts a cosmological evolution of the BTFR normalization as the mass in the observable Universe evolves. We test these predictions against recent high-redshift observations ($z \sim 2$ and $z \sim 5$), finding that the data 
 favor a flat, baryon-only, $\Lambda$-dominated quasi-de Sitter universe ($\Omega_m \approx 0.05$, $\Omega_\Lambda \approx 0.95$) over standard MOND or $\Lambda$CDM parameters. These results suggest that flat galactic rotation curves may be interpreted as compelling evidence in favor of Mach's Principle, in at least some of its versions.}
\end{abstract}
\section{Introduction}

For half a century, since the discovery of asymptotically flat rotation curves in disk galaxies (Rubin et al. 1978), evidence has accumulated for the presence of mass-energy discrepancies in the Universe. With the discovery in the late 1990s of the accelerating expansion of the Universe observed in distant supernovae (Perlmutter et al. 1999), the cosmological constant $\Lambda$-Cold Dark Matter ($\Lambda$CDM) paradigm has emerged as the concordance model in cosmology. The great agreement of independent cosmological observables from Gpc to Mpc scales lends an air of inevitability to the $\Lambda$CDM model; nevertheless, this success relies partially on extra assumptions such as the baryonic fraction, which varies considerably (Famaey \& McGaugh 2012), behaving almost as a free parameter nowadays. In addition, one of the most important tests of $\Lambda$CDM as a scientific hypothesis is positive laboratory evidence for the existence of dark matter; however, searches for WIMPs (and alternatives) are now rather mature and not particularly encouraging. Theorists have postulated ever more extreme properties for dark matter particles instead of searching for alternatives (Merritt 2017).

Through these years, there has also been a large variety of attempts to alter the relevant physical laws in order to remove the need for cold dark matter, with 
 one survivor: MOdified Newtonian Dynamics (MOND), which proposes that the law of inertia or gravity takes on a specific non-Newtonian form at accelerations well below a definite universal value $a_0 \approx 1.2 \times 10^{-10}\,{\rm m/s}^2$ (Milgrom 1983). MONDian dynamics works very well on the scale of galaxies, with results that can be summarized in 10 `Kepler-like' empirical laws of galactic motions (Famaey \& McGaugh 2012), including those named as the `Mass-Discrepancy Acceleration' (Sanders 1990; McGaugh 2004) and the `Baryonic Tully-Fisher' (Lelli et al. 2016) relations. On those scales, MOND precisely predicts the force in galaxies from the observed distribution of baryonic matter (without the need to hypothesize any `dark sector'), being quite problematic for CDM since it would require an intimate dark matter-baryon coupling. However, MOND's success appears to break down at the scales of clusters of galaxies, still requiring an extra factor of 2--3 of unobserved matter (Famaey \& McGaugh 2012).

However, this piece of evidence associated with the MOND paradigm has been mostly ignored by the scientific community. In fact, Merritt (2017) documented this statement through analyzing graduate-level texts on cosmology and galaxy formation, in particular, how the failure of laboratory detection for dark matter is generally not discussed, even less a universal acceleration scale $a_0$ or some of its most relevant predictions such as the mass discrepancy-acceleration relation. One of the reasons is probably that the success of astrophysics in the last two centuries fundamentally relies on the success of physics, particularly on its proper formulation that allowed it to be applied in regimes far from those laboratory-tested (something that is unfortunately missed in other areas of knowledge, see e.g., Escala 2019). This success has made the mainstream astronomical community very reluctant to explore alternatives where physical equations would be modified. Nevertheless, the mere need to conjecture the existence of CDM and the cosmological constant $\Lambda$ implies that something must be added or modified within established physics, either in particle physics (CDM), gravity/inertia (MOND), or in cosmology itself ($\Lambda$ and alternatives). Each of these three alternatives is a major change in our understanding of the physical world.

In this paper, we will explore possibilities within the subset of the latter option, looking for alternatives that do not fully rely on the MOND or $\Lambda$CDM paradigms; in particular, we will study the possibility of an interpretation based on Mach's principle for the phenomena associated with asymptotically flat rotation curves in galaxies. We start with an analysis of the implications of having $a_0$ as a new universal constant in \S 2, as is usually stated in MOND. We continue in \S 3, proposing that the MOND paradigm could be interpreted in terms of Mach's principle, also rewriting the Baryonic Tully-Fisher relation in a cosmological format. In \S 4, we study the cosmological implications of a Machian interpretation, contrasting predictions for the Baryonic Tully-Fisher relation against recent observational data, with promising results. Finally, we discuss the results and their implications in \S 5.

\section{MOND's acceleration $a_0$ as a fundamental constant}

Planck's constant, the velocity of light, and Newton's constant of gravitation are considered three dimensional constants more fundamental than any other physical constants, which determined the three basic units first identified more than a century ago by Planck (1899), namely the Planck units: length $L_P$, mass $M_P$, and time $T_P$. The motivation was that in quantum mechanics, there was a minimum quantum of action given by Planck's constant $\hbar$; in relativity, there was a maximum velocity given by the velocity of light $c$; and in classical gravity, the strength of the force between two objects was determined by Newton's constant of gravitation $G$. This was not the first attempt at a natural system of units (e.g., Stoney in 1881 used the electric charge before the discovery of a minimum quantum of action $\hbar$), but up to date, Planck units are still considered a consensus among physicists as the natural system of units (see, however, discussions by Duff et al. 2002 or Duff 2015 for a different view). 
 
Now, if the aim is to include $a_0 = 1.2 \times 10^{-10}\,{\rm m/s}^2$ in the list of fundamental constants, it is necessary to replace one constant, as Planck did with the electric charge of Stoney's units using his quantum constant $\hbar$. Since MOND introduces a new constant, $a_0$, with the dimensions of acceleration that marks the boundary between the validity domains of Newtonian and MONDian dynamics, whose significance is similar to that of $\hbar$ in the quantum context, or to that of the speed of light $c$ in the relativistic context (Milgrom 1983, 1999), $a_0$ should naturally replace the constant of gravitation $G$, since no equivalent role has been found for $G$ yet. Moreover, $G$ is the only one of the three constants not used for the definition of the International System of Units, probably due to the lack of accuracy in its determination.

Performing dimensional analysis on $a_0$, $c$, and $\hbar$, we arrive at a first set of MONDian units: $T_M=\frac{c}{a_0}=2.5 \times 10^{18} \,{\rm s}$, $L_M=\frac{c^2}{a_0}=7.5 \times 10^{26}\, {\rm m}$, $M_M=\frac{\hbar a_0}{c^3}= 4.7 \times 10^{-70}\,{\rm kg}$, the first two being of cosmological significance (i.e., similar to the Hubble time and radius in the observable Universe) and the third one, $M_M$, being too small to have any physical significance. Another possibility is to use $G a_0$, which appears in several relations such as the Baryonic Tully-Fisher relation, giving again a mixture of extremely small ($L_{M}=\frac{\hbar G a_0}{c^5} \approx 10^{-97}\,{\rm m}$ and $T_{M} = \frac{\hbar G a_0}{c^6} \approx 10^{-105}\,{\rm s}$) and cosmological scales ($M_{M} = \frac{c^4}{G a_0} \approx 10^{54} \,{\rm kg} \sim M_u$). Only if $a_0/G$, $\hbar$, and $c$ are chosen does it give units of the same (tiny) scale, but $a_0/G$ has a less motivated relevance (related to inner stability in galaxies, Famaey \& McGaugh 2012, thus far from being considered fundamental) and the related scales have an unclear physical meaning again.

Therefore, it is very hard to argue in favor of replacing $G$ with $a_0$ in the list of most fundamental physical constants (with or without a combination with $G$). On the other hand, if we choose $a_0$, $c$, and $G$, we derive for $T_M$, $L_M$, and $M_M$ the fundamental scales of the observable Universe:
\begin{equation}
(a) \,\,T_M= \frac{c}{a_0}  \sim \, t_u
 \,\, \,\, \,\,\,\,(b)\,\,  L_M=\frac{c^2}{a_0}  \sim R_u\,\,\,\,\,\,\,\,(c)\,\, M_M=\frac{c^4}{G a_0}
\sim M_u \,\,\, ,
\label{Units}
\end{equation}
which are approximately the cosmologically preferred values for the current Universe, for the Hubble time $t_u$, radius $R_u$, and total mass $M_u$ in the Universe. In particular, for the mass $M_M=\frac{c^4}{G a_0} = 1.01 \times 10^{54}$ kg, the $M_M$ value is strikingly similar to the one required by the total mass-energy in the observable Universe to have a flat geometry ($ M_u \approx 1.5 \times 10^{54}$ kg; Planck Collaboration, Aghanim et al. 2020). 

These relations can be interpreted as a version of the coincidence $a_0 \approx c H_0$, noticed as early as the development of MOND (Milgrom 1983), a coincidence that has also been expanded to a relation with the cosmological constant $\Lambda$ ($a_0 \approx c^2 \sqrt{\Lambda}$; Milgrom 1999). However, in a $\Lambda$ (dark energy) dominated state like the current state of the Universe (Planck Collaboration, Aghanim et al. 2020), the Friedmann equation in a flat universe with a cosmological constant already predicts $H_0 \approx c\sqrt{\Lambda/3}$; therefore, this second coincidence can just be derived from the first one (or vice versa). On the other hand, the relations in Eq. \ref{Units} can also be added to the series of coincidences that have puzzled physicists for decades ever since Dirac's Large Number Hypothesis (Dirac 1937), which has also motivated alternative theories of gravity like Jordan (1952), Brans \& Dicke (1961), Hoyle \& Narlikar (1964), and others (see Peebles 2017 for an historical review).

In summary, it is hard to argue that $G$ should be replaced by $a_0$, and therefore we can conclude that $a_0$ is probably not a new fundamental constant with the same status as $\hbar$ and $c$ to define a definite set of natural base units that will replace Planck's ones, as is generally stated for MONDian dynamics (Milgrom 2015). Nevertheless, the role of $a_0$ (and MOND) can still be viewed as a useful prescription to describe the dynamics of galactic motions. Throughout this work, MOND will be used as an algorithm for calculating the distribution of force in an astronomical object from the observed distribution of baryonic matter (i.e., the minimalist definition of MOND, according to Sanders 2008).  

In the next section, we will argue in favor of a Machian interpretation for the `apparent coincidence' between $a_0$ and cosmological parameters in Eq. \ref{Units}. For example, a simple interpretation would be that $a_0 \approx c^2/R_u \approx c^4/GM_u \approx c/t_u$ suggests that there is a moment when the acceleration in a galaxy becomes so small that it eventually `notices' the boundary of the Universe and its dynamics change. We will expand in the next section on these interpretations, after rewriting some empirical results in a cosmologically suggestive form.

\section{Baryonic Tully-Fisher Relation Rewritten in a Cosmological Form and Its Relation with Mach's Principle} 

Mach's Principle---the idea that the local inertia of an object is determined by the distribution of all the matter in the rest of the Universe---remains one of the most provocative concepts in physics, probably because just as motion can be considered meaningful only relative to the rest of the matter in the Universe, it is logical to conjecture that inertia is also meaningful only relative to the rest of the Universe. This principle highly influenced Einstein in the development of General Relativity (who also coined the term; Dicke 1959), but was ultimately not explicitly included. Instead, Einstein's Equivalence Principle gives an answer to the problem of inertia that does not refer to Newton's absolute space, yet does not include Mach's Principle; therefore, it is considered an open issue (Weinberg 1972). A possible solution that can be found is a cosmological solution within General Relativity with boundary conditions that fulfill Mach's conclusions (Dicke 1959). A minority considers that Einstein's Cosmological Principle already makes it Machian (Peebles 2017).

Since the relations in Eq. \ref{Units} are clearly Machian in spirit, they serve as motivation to rewrite some of the results describing galaxy rotation curves in a cosmologically suggestive form. Also, since the relations in Eq. \ref{Units} include the characteristic acceleration $a_0$, it is easier to start from some MOND results, for example, the derivation of the Baryonic Tully-Fisher relation, which is a prediction within MONDian dynamics. Starting with the (deep-)MOND relation for low accelerations, $g_{\rm obs} = \sqrt{g_{\rm bar} \cdot a_0}$, when substituting $g_{\rm obs} = \frac{V^2}{R}$ (centripetal acceleration) and $g_{\rm bar} = \frac{G M_{\rm bar}}{R^2}$, the $R$ terms cancel out, leaving a Baryonic Tully-Fisher relation in the deep MOND limit: $V^4 = G M_{\rm bar} a_0$ (Famaey \& McGaugh 2012). Assuming the link between the acceleration constant $a_0$ and the total mass of the Universe $M_u$, i.e., $G a_0 = c^4/M_u$ from Eq. \ref{Units}c, leads to the following simple (but very suggestive) expression of the Baryonic Tully-Fisher relation for low accelerations:
\begin{equation}
\frac{M_{\rm bar}}{M_{u}} = \left( \frac{V}{c} \right)^4   \,\,\, ,
\label{TF}
\end{equation}
which has a straightforward cosmological interpretation and a validity that is not restricted to MONDian dynamics, since it can be considered empirically valid (in fact, in \S 3.1 we discuss some MOND-independent derivations).

\begin{figure}[h!]
\begin{center}
\includegraphics[width=11.9cm]{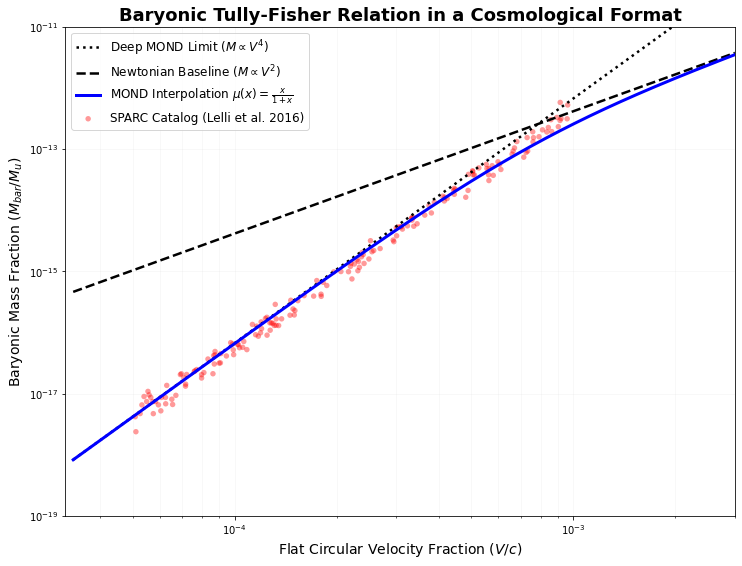}
\caption{The Baryonic Tully-Fisher relation rewritten in the cosmologically suggestive form proposed in this work. The red circles display the SPARC dataset of galaxies compiled by Lelli et al. (2016). The blue curve represents the standard MOND interpolating function, $\mu(x) = \frac{x}{1+x}$, translated into physical parameters with a characteristic transition radius $R_t = 15$ kpc. The black dashed line indicates the Newtonian regime ($M_{\rm bar} \propto V^2$) at high velocities, while the black dotted line represents the deep MOND limit ($M_{\rm bar} \propto V^4$) at low velocities.}
\label{F1}
\end{center}
\end{figure}

Figure \ref{F1} shows the Baryonic Tully-Fisher relation rewritten in a cosmologically suggestive form. Red circles display the SPARC (Spitzer Photometry and Accurate Rotation Curves) dataset of galaxies, compiled by Lelli, McGaugh, and Schombert (2016). The blue curve uses the standard MOND interpolating function, $\mu(x) = \frac{x}{1+x}$, expressed in terms of the acceleration ratio $x = g/a_0$, translated into physical parameters by evaluating $g$ at a characteristic transition radius $R_t$, where $g = V^2/R_t$. This interpolating function yields the following explicit formula for the baryonic mass: $M_{\rm bar}(V) = \frac{V^4 R_t}{G(a_0 R_t + V^2)}$, which leads to the Newtonian regime ($V^2 \gg a_0 R_t$) at high velocities, yielding the formula $M_{\rm bar} \approx \frac{V^2 R_t}{G}$ valid for a virialized system under Newtonian gravity (black dashed line), while at low velocities ($V^2 \ll a_0 R_t$) it leads to the classic Baryonic Tully-Fisher relation in the deep MOND regime $M_{\rm bar} \approx \frac{V^4}{G a_0}$ (black dotted line). We see an overall good agreement of the blue curve with SPARC data (for $R_t = 15 \, \rm kpc$), as expected since it is basically the same Baryonic Tully-Fisher  relation, only rewritten in a cosmologically suggestive form. 

It is straightforward to see that under this interpretation, the Newtonian limit ($M\propto V^2$) arises at high enough accelerations where local dynamics fully dominate over the cosmological boundary, while the deep MOND regime appears when motions are dominated by the cosmological boundary. In this framework, $a_0$ is not a fundamental constant of nature like the speed of light; it is only a scaling constant that tells us when a local system has accelerations slow enough to finally `feel the edge of the Universe'. In the next subsection, we see some examples of how this `feeling the edge of the Universe' could be realized.

\subsection{A couple of examples for Machian derivations}

We will discuss some examples of Machian theories that predict a Tully-Fisher relation that can be expressed like Eq. \ref{TF}, not aiming to claim support in favor of a particular theory, since some are already considered invalid (those that imply cosmological variations in $G$) or are still considered too exotic by most physicists. Instead, our aim is to illustrate that any Machian derivation should give a relation like Eq. \ref{TF}, partially because the Mach Principle is currently  a  loosely specified idea, with several possible definitions (at least 10, according to Bondi \& Samuel 1996). The modern Machian view is basically  only restricted  to something global,  at universe scales, that somewhat affects   local dynamics.  

For simplicity, we start with Sciama's inertial induction, which proposed that a particle's inertia is actually a gravitational inductive effect, through a mechanism similar to the one seen in Maxwell's equations for electromagnetic induction (Sciama 1953). In Sciama's theory, the relationship between the gravitational constant $G$, the mass of the observable Universe $M_u$, and its radius $R_u$ is defined by (Sciama 1953): 
\begin{equation}
\frac{GM_u}{R_uc^2} \approx 1 \,\,\,  . 
\label{Sciama}
\end{equation}
If we consider the `Machian' acceleration to be the characteristic gravitational acceleration produced by the mass of the Universe at its own boundary, the Hubble scale, we can define a `cosmic acceleration': $a_c = GM_u/R_u^2$. When we substitute Sciama's definition of $G$ (from Eq. \ref{Sciama}) into this acceleration, the mass $M_u$ cancels out, leaving $a_c = c^2/R_u$, which is the relation fulfilled in Eq. \ref{Units}b by the MONDian acceleration constant $a_0$ and therefore, also equivalent to Eq. \ref{TF}. Moreover, in the context of Sciama's inertial induction it can be shown (Gin\'e 2009) that this leads to the same MONDian interpolation function $\mu(x) = \frac{x}{1+x}$ used in Figure \ref{F1}, as well as when applying just the relativity principle of motion with cosmic acceleration (Gin\'e 2012).   
 
Nevertheless, it is straightforward that Eq. \ref{Sciama} implies a cosmological variation for the constant of gravitation $G$, since $R_u$ and $M_u$ vary across cosmological time (while $c$ is constant). The same holds for other theories that follow up on this idea, such as the Brans-Dicke theory (Brans \& Dicke 1961) and scalar-tensor theories of gravity in general, or other results like Reissner (1915) and Schr\"odinger (1925), which arrived at the same conclusion in Eq. \ref{Sciama}. Since modern measurements restrict $G$ variations to a level of $\dot{G}/G < 10^{-12}$ (Peebles 2017), it is not worthy to discuss in further detail these Machian theories in which inertia is derived from gravity, in particular, where $G$ is derived from the parameters of the Universe. This conclusion is in concordance with the results in \S 2, where $G$ cannot be replaced (yet) as a fundamental constant, at least by MOND's acceleration $a_0$.

Another set of derivations includes those related to possible interpretations of the cosmological constant $\Lambda$, for example, those that relate Unruh radiation to a minimum acceleration constant $a_0$ (McCulloch 2007, 2013). According to the Unruh effect, an object accelerating at a given rate does not experience `empty' space; instead, it perceives the vacuum as being filled with a thermal bath of radiation (Unruh radiation; Unruh 1976). At the same time, the Universe itself has a `temperature' associated with its boundary, the cosmological horizon, and this Hawking radiation onto the cosmological horizon is called the Gibbons-Hawking effect (Gibbons \& Hawking 1977). The basis of this process lies in Unruh radiation wavelengths truncated by the cosmological horizon, since at very low accelerations ($a < a_0$), these waves become so long that they can be truncated by the cosmic horizon (McCulloch 2007).

Under this context of a cosmological constant $\Lambda$ modeled as a quantum bath, to find the point where the new behavior begins, we look for the threshold where local thermal effects (from moving) are equal to global thermal effects (from the Universe itself, thus Machian in spirit). As you accelerate, the vacuum appears `warm' to you (local term), where the Unruh temperature $T_U$ perceived is $T_U = \frac{\hbar a_U}{2\pi c k_B}$, relating the acceleration $a_U$ to this temperature. In the Gibbons-Hawking temperature $T_{\rm GH}$, the Universe is expanding and has a cosmological horizon (global term); this horizon also has a temperature, the Gibbons-Hawking temperature: $T_{\rm GH} = \frac{\hbar c}{2\pi k_B R_u}$, where $R_u$ is the observable Universe radius or cosmic horizon (i.e., $R_u = c/H_0$). Since the Unruh temperature has a similar form to the Gibbons-Hawking temperature, when we set the two temperatures equal to find the transition acceleration,
\begin{equation}
T_U = \frac{\hbar a_U}{2\pi c k_B} = \frac{\hbar c}{2\pi k_B R_u}= T_{\rm GH}  \,\,\, ,
\label{UnruhHG}
\end{equation}
most terms cancel out, particularly the quantum ($\hbar$) and thermodynamic ($k_B$) ones, leaving again $a_U = c^2/R_u$, the relation fulfilled in Eq. \ref{Units}b by the acceleration constant $a_0$, therefore also fulfilling a Baryonic Tully-Fisher relation equivalent to Eq. \ref{TF}.

In summary, using two examples with completely different physics and assumptions, we illustrated that any Machian derivation should give a Baryonic Tully-Fisher relation like Eq. \ref{TF}, as long as it relies on a minimum (or base) acceleration that is linked to some cosmological parameter (i.e., $R_u$ or $M_u$). This is simply because, in order to be dimensionally correct, it must fulfill relations like those listed in Eq. \ref{Units}.

\section{Cosmological Implications for the Baryonic Tully-Fisher relation}

In the previous section, we concluded that regardless of the particular interpretation, any Machian derivation must fulfill a relation of the type $a_0(z) \approx c^2/R_u(z)$. This predicts cosmological variations in the Baryonic Tully-Fisher relation, Eq. \ref{TF}, through changes in the total mass of the Universe $M_u$ as follows:
\begin{equation}
M_u (z) = \frac{c^4}{G a_0(z)} 
 \,\,\, ,
\label{a0(z)}
\end{equation}
which depends on the different density parameters ($\Omega_m, \Omega_\Lambda, \Omega_k$) as 

$a_0(z) = a_0 \sqrt{\Omega_m(1+z)^3 + \Omega_k(1+z)^2 + \Omega_\Lambda}$, where $\Omega_m$ is the matter density parameter (baryons and/or dark matter), $\Omega_\Lambda$ is the dark energy (cosmological constant) parameter, $\Omega_k = 1 - (\Omega_m + \Omega_\Lambda)$ is the curvature parameter, and $a_0 
= 1.2 \times 10^{-10} \, {\rm m/s}^2$ is the local value today. This directly affects the normalization in the Baryonic Tully-Fisher relation (i.e., $M_u (z) \, c^{-4} = 1/Ga_0(z)$ in Eq. \ref{TF}), predicting that the relation must evolve on cosmological timescales. Testing the evolution of $M_u (z)$ is probably the strongest constraint for the possible role of Mach's Principle in explaining flat galactic rotation curves, much more so than any partial success of a particular Machian theory at this stage.

Genzel et al. (2017, 2020), later extended by Nestor-Shachar et al. (2023), reported (in the context of the CDM scenario) extreme cosmological variations of dark matter fractions, in particular, the existence of strongly baryon-dominated disk galaxies at $z \sim 2$, where flat rotation curves are not typically present (Genzel et al. 2017, 2020). Milgrom (2017) argued, in the context of the MONDian paradigm, that the `falling' rotation curves in Genzel et al. (2017) occur because MOND effects had not `taken over' yet, since dynamical accelerations in such $z \sim 2$ samples are higher compared to ones in the nearby Universe (i.e., accelerations at half-light radii, $g(R_{1/2}) = [3-11]\,a_0$). Milgrom (2017) also argued that this directly challenges the idea that $a_0$ cosmologically scales upward (proportional to $(1+z)^{3/2}$), since it implies $a_0(z) \sim 5 \, a_0$ at $z \sim 2$, already in the acceleration regime of the Genzel et al. (2017) sample ($[3-11]\,a_0$).

However, in general, the acceleration threshold should cosmologically vary in a Machian context according to $a_0(z) = a_0 \sqrt{\Omega_m(1+z)^3 + \Omega_k(1+z)^2 + \Omega_\Lambda}$, where $\Omega_k = 1 - (\Omega_m + \Omega_\Lambda) = 0$ for a flat Universe (Planck Collaboration, Aghanim et al. 2020); only in a matter-only flat universe ($\Omega_m = 1, \Omega_\Lambda = 0$) does $a_0(z)$ scale exactly as $(1+z)^{3/2}$, as stated in Milgrom (2017). Since there is no observational evidence that currently favors such an $\Omega_m = 1$ universe (moreover, assuming the total absence of dark matter in the Universe, as in Milgrom 2017), a more realistic estimation would be to assume a flat Universe (strongly supported by CMB observations; Planck Collaboration, Aghanim et al. 2020) with a more standard $\Omega_m = 0.05$ (due to baryonic matter alone) and $\Omega_\Lambda = 1 - \Omega_m = 0.95$, which gives $a_0(z) \sim 1.5\,a_0$ at $z \sim 2$, therefore compatible with the `falling' rotation curves seen in the Genzel et al. (2017) sample (i.e., less than $g(R_{1/2}) = [3-11]\,a_0$).

One more quantitative possibility is to contrast fitted normalizations in the Baryonic Tully-Fisher relation with the prediction from Eq. \ref{a0(z)}. Since the reported values are on a log scale, Eq. \ref{TF} should be rewritten as $\log_{10} M_{\rm bar} = 4 \log_{10} \left( \frac{V_{\rm circ}}{c} \right) + \log_{10} M_u(z)$; therefore, Eq. \ref{a0(z)} predicts that any offset in mass at a fixed circular velocity ($V_{\rm circ}$) is given by:
 \begin{equation}
 \Delta \log_{10} M_{\rm bar} =  \log_{10}  \left( \frac{M_u(z)}{M_u(0)} \right) 
= -\log_{10} \sqrt{\Omega_m (1+z)^3 + \Omega_k (1+z)^2 + \Omega_\Lambda}   \,\, .
\label{Delta}
\end{equation}

\begin{figure}[h!]
\begin{center}
\includegraphics[width=11.9cm]{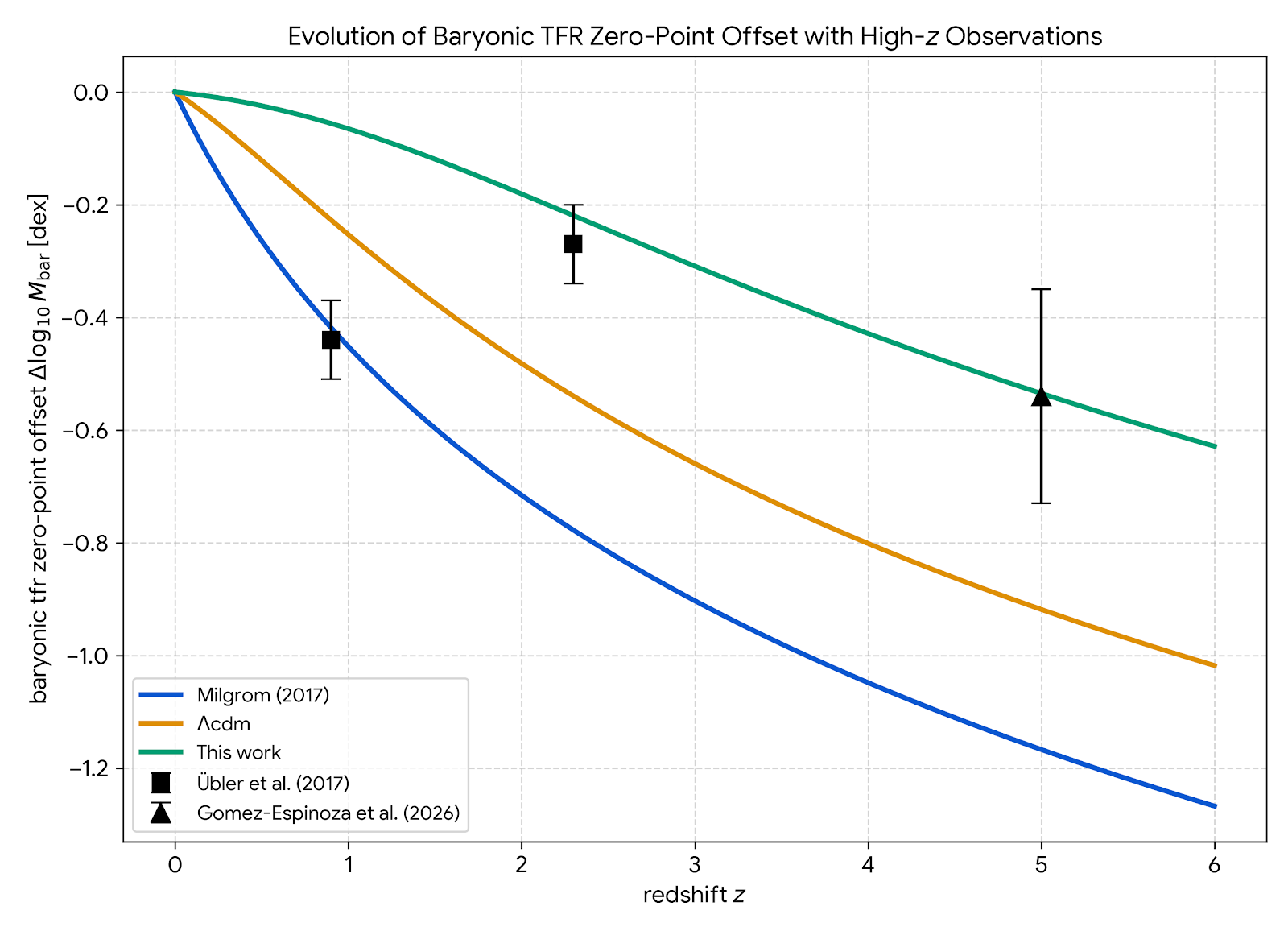}
\caption{Cosmological evolution of the vertical zero-point ($\Delta M_{\rm bar}$) in the Baryonic Tully-Fisher relation. The symbols denote high-$z$ observations that are compared with the prediction $a_0 \propto (1+z)^{3/2}$ in Milgrom (2017), the one expected from standard parameters in $\Lambda$CDM cosmology ($\Omega_m = 0.315$, $\Omega_\Lambda = 0.685$), and with the cosmological parameters used in this work ($\Omega_m = 0.05$, $\Omega_\Lambda = 0.95$). In all cases, a flat Universe is assumed as expected from CMB observations (Planck Collaboration, Aghanim et al. 2020).}  
\label{F2}
\end{center}
\end{figure}

In Fig. \ref{F2}, we perform this comparison of the prediction from Eq. \ref{Delta} for different cosmologies and fitted normalizations in the Baryonic Tully-Fisher relation at different redshifts. The zero-point mass offsets at a fixed circular velocity are denoted by square (\"Ubler et al. 2017) and triangle (G\'omez-Espinoza et al. 2026) symbols in Fig. \ref{F2}. A significant cosmological variation is indeed observed for the fitted normalizations in the Baryonic Tully-Fisher relation, something inconsistent with standard MOND (where it should be fixed equal to $G a_0$), but expected in a Machian interpretation with $M_u(z)$. The blue curve is the prediction implemented in Milgrom (2017) (baryonic matter-only Universe, $\Omega_m = 1, \Omega_\Lambda = 0$), the yellow line is the standard $\Lambda$CDM cosmology ($\Omega_m = 0.315, \Omega_\Lambda = 0.685$), and the green one represents the cosmological parameters used in this work, both compatible with Planck's flat Universe and the absence of dark matter in the Machian interpretation, thus close to a de Sitter universe ($\Omega_m = 0.05, \Omega_\Lambda = 0.95$).

We find that at higher redshifts ($z \sim 2$ and $5$), data points clearly favor the cosmology consistent with this work, i.e., a quasi-de Sitter universe without dark matter, with the lower redshift point ($z = 0.9$) being consistent only with the flat baryon-only universe (standard $\Lambda$CDM cosmology parameters are not consistent with any of the points). Nevertheless, $z = 0.9$ datapoint is also not consistent with the updated model (based on CDM) in G\'omez-Espinoza et al. (2026), which is based on dark matter fraction ($f_{\rm DM}$) estimations determined in the RC100 sample (Nestor-Shachar et al. 2023); therefore, $z = 0.9$ datapoint is is also not consistent with an independent measurement within the context of the CDM paradigm. This also illustrates how $f_{\rm DM}$ operates nowadays almost as an extra free parameter, without any reference to why it should vary on cosmological timescales, as long as it is convenient to fit the data under study.

In summary, we find that a significant cosmological variation is indeed observed for the fitted normalizations in the Baryonic Tully-Fisher relation, as predicted in the Machian interpretation. Also, higher-redshift ($z \sim 2$ and $5$) data points clearly favor the cosmology currently consistent with this Machian scenario (no dark matter in a $\Lambda$-dominated flat universe). Moreover, there is considerable evidence at higher redshifts for the existence of strongly baryon-dominated disk galaxies (Genzel et al. 2017, 2020; Nestor-Shachar et al. 2023), challenging CDM but in agreement with the Machian interpretation for the MOND paradigm (with accelerations still higher than $a_0(z)$). Unfortunately, it is hard to advance toward a consensus if those works systematically ignore the MOND paradigm, not even referring to a characteristic acceleration $a_0$. A similar case applies to those studying Baryonic Tully-Fisher relation variations (e.g., \"Ubler et al. 2017), again not referring at all to $a_0$, despite such a relation originating as a prediction of MOND that has been extensively studied within the MONDian context (e.g., Lelli et al. 2016).

\section{Discussion} 

In this paper, we have explored a novel cosmological framework where asymptotically flat galactic rotation curves are interpreted not through the lens of hypothetical dark matter particles or direct modifications to fundamental physical laws, but as a consequence of Mach's Principle. By highlighting that the MOND acceleration scale $a_0$ matches the fundamental scales of the observable Universe ($a_0 \approx c^4/GM_u \approx c^2/R_u \approx c/t_u$), we argue that the deep MOND regime represents a threshold where local dynamics begin to feel the global boundary conditions of the cosmos.

A critical achievement of this approach is the successful prediction of the cosmological evolution of the Baryonic Tully-Fisher relation's zero-point. When evaluated against high-redshift data up to $z \sim 5$, the observed offsets significantly favor a flat, baryon-only, $\Lambda$-dominated universe ($\Omega_m \approx 0.05, \Omega_\Lambda \approx 0.95$). While this shifts the cosmic composition from the standard $\Lambda$CDM values ($\Omega_m \sim 0.3$), reducing the matter contribution to a negligible fraction ($<5\%$) transforms the Universe into a quasi-de Sitter state. This aligns closely with specific definitions of Mach's Principle, where global geometry and mass distribution directly constrain local inertial states (e.g., $\Omega$ is a definite number of order unity; Mach8 in Bondi \& Samuel 1996).

The empirical success of MOND as a predictive algorithm on galactic scales, combined with the observed evolution of the BTFR normalization at high redshifts, motivates a shift away from the dark matter paradigm. While $\Lambda$CDM relies heavily on the dark matter fraction ($f_{\rm DM}$) operating essentially as a free parameter to reconcile galactic dynamics across epochs and scales, which is critical to its success, the Machian interpretation offers a predictive, parameter-free explanation for the coupling between baryonic matter and cosmic scales. Future high-redshift observations will be vital to further constrain the evolution of the BTFR zero-point and test the boundaries of this Machian perspective.

Beyond individual galactic scales, this Machian framework offers a fresh perspective on larger cosmological structures, specifically regarding Baryon Acoustic Oscillations (BAO) and galaxy clusters. While global cosmological observables like the cosmic microwave background and BAO have historically presented severe challenges to dark-matter-free models, there are enough uncertainties that there is still room for the BAO second peak to be successfully reproduced within an effective MONDian gravity framework, especially without knowing the specific Machian process that operates. Within our interpretation, this would imply that the early Universe's plasma dynamics were already operating under the influence of an effective acceleration threshold $a_0(z)$ scaled to the boundary conditions of the nascent cosmos.

Another well-known breakdown of standard MOND is on the scale of galaxy clusters, where it typically falls short by a factor of 2 or 3 of unobserved matter; this could be reconciled by recognizing that clusters are inherently too complex to serve as clean, dynamically isolated systems. The thermodynamic complexity of the intracluster medium could in principle mask the local-to-global acceleration transitions, for example compared to the case of an Unruh acceleration with respect to the vacuum, in which case the cluster discrepancy arises from incomplete baryonic accounting rather than a failure of the global boundary condition. A similar case applies to gravitational lensing measurements, which in principle could also depend on the cosmological boundary. In addition, there are claims that galaxy clusters could fulfill some MONDian empirical relations such as the `Mass-Discrepancy Acceleration' relation, but with a higher $a_0$ (Tian et al. 2020), as expected from a Machian viewpoint.

Finally, it is worth emphasizing a profound conceptual contrast that emerges from this perspective regarding the missing mass problem. While the standard $\Lambda$CDM paradigm posits that the Universe is filled with an enormous, dominant mass of dark matter, the raw observational fact is remarkably subtle: a small, systematic velocity anomaly occurring primarily at the remote outskirts of galaxies, which never appears to decay so far. Rather than requiring vast, invisible halos of exotic particles to account for this peripheral behavior, these effects can be naturally reinterpreted without invoking missing matter. Instead, this edge-effect can be viewed as the manifestation of a global boundary condition imposed on local systems by the Universe itself. Under this Machian viewpoint, the characteristic acceleration scale $a_0(z)$ represents the threshold where local Newtonian dynamics smoothly interface with the causal boundary of the cosmic horizon, transforming a complex dark matter inventory problem into a fundamental property of cosmic structure and also, solving the long-standing problem of incorporating Mach's conclusions into our worldview. We conclude with a quote from Einstein (1922), which may be translated as: `It is contrary to the scientific mode of understanding to postulate a thing which acts, but which cannot be acted upon.' Sometimes CDM seems a prime example of such a `thing'.

I thank Juan Molina for helpful discussions on the cosmological evolution of the Tully-Fisher relation within the $\Lambda$CDM framework. I acknowledge financial support from BASAL grant AFB-170002.

\end{document}